\documentclass[twocolumn]{aastex63}
\usepackage{natbib}
\usepackage{amsmath}
\usepackage{chngcntr}
\usepackage{hyperref}
\usepackage[figuresright]{rotating}
\usepackage{ctable}
\usepackage{graphicx}
\usepackage{realboxes}
\usepackage{epsfig}
\usepackage{threeparttablex, tablefootnote}
\usepackage{subfigure}
\usepackage{graphicx}
\newcommand{\uat}[2]{\href{http://astrothesaurus.org/uat/#2}{#1 (#2)}}
\usepackage{CJK}
\shortauthors{Ding et al.}
\usepackage{threeparttable}
\begin{document}

\begin{CJK*}{UTF8}{gbsn}
%\title{Chromospheric activities across the Hertzsprung-Russell diagram: evidence of enhanced binary $S$ index}

%\title{Complex behaviour of chromospheric activity in binaries compared to single stars}

\title{Double-edged sword: the influence of tidal interaction on stellar activity in binaries}

%\title{Activity comparison between binaries and single stars}

\author{Yuedan Ding}
\affiliation{Department of Physics, Hebei Normal University, Shijiazhuang 050024, People's Republic of China}
\affiliation{Guo Shoujing Institute for Astronomy, Hebei Normal University, Shijiazhuang 050024, People's Republic of China}

\author{Shidi Zhang}
\affiliation{Department of Physics, Hebei Normal University, Shijiazhuang 050024, People's Republic of China}
\affiliation{Guo Shoujing Institute for Astronomy, Hebei Normal University, Shijiazhuang 050024, People's Republic of China}
\affil{Co-first author}

\author{Henggeng Han}
\affil{Key Laboratory of Optical Astronomy, National Astronomical Observatories, Chinese Academy of Sciences, Beijing 100101, People's Republic of China}

\author[0000-0003-1359-9908]{Wenyuan Cui}
\affiliation{Department of Physics, Hebei Normal University, Shijiazhuang 050024, People's Republic of China}
\affiliation{Guo Shoujing Institute for Astronomy, Hebei Normal University, Shijiazhuang 050024, People's Republic of China}

\author[0000-0003-3116-5038]{Song Wang}
\affiliation{Key Laboratory of Optical Astronomy, National Astronomical Observatories, Chinese Academy of Sciences, Beijing 100101, People's Republic of China}
\affiliation{Institute for Frontiers in Astronomy and Astrophysics, Beijing Normal University, Beijing 102206, China}

\author{Min Fang}
\affiliation{Purple Mountain Observatory, Chinese Academy of Sciences, 10 Yuanhua Road, Nanjing 210023,  People's Republic of China;}

\author{Yawei Gao}
\affiliation{Department of Physics, Hebei Normal University, Shijiazhuang 050024, People's Republic of China}
\affiliation{Guo Shoujing Institute for Astronomy, Hebei Normal University, Shijiazhuang 050024, People's Republic of China}

\correspondingauthor{Wenyuan Cui; Song Wang}
\email{wenyuancui@126.com; songw@bao.ac.cn}

\begin{abstract}

Using the LAMOST DR7 low-resolution spectra, we carried out a systematic study of stellar chromospheric activity in both single and binary stars.
We constructed a binary sample and a single-star sample, mainly using the binary belt and the main sequence in the Hertzsprung-Russell diagram, respectively.
By comparing the $S$ indices between single and binary stars within each color bin, we found for K type stars, binaries exhibit enhanced activity compared to single stars, which could be attributed to the increase in spin rate caused by tidal synchronization or to the interactions of magnetic fields.
Both single stars and binaries fall on a common sequence in the activity-period relation, indicating that chromospheric activities of binaries are dominated by the more active components.
More intriguingly, in some color ranges, a slight decline of the $S$ index for smaller orbital period was observed for binary stars.
Although the possibility of sample selection effects cannot be excluded, this may mark the first example of super-saturation (i.e., caused by reduced active regions) being detected in chromospheric activity, or provide evidence of the suppressing effect on the magnetic dynamo and stellar activities by strong tidal interaction in very close binaries.
Our study suggests that tidal interaction acts as a double-edged sword in relation to stellar activities.

\end{abstract}

%\keywords{stars: late-type -- stars: activity -- stars:chromospheres}

\section{Introduction} 

Stellar activities are external manifestation of the induction and relaxation of magnetic fields.
According to the dynamo theories, magnetic fields are generated by stellar rotation and convection.
In the $\alpha$--$\Omega$ dynamo, magnetic fields primarily arise from differential rotation at the tachocline \citep{1955ApJ...122..293P, 1955ApJ...121..491P, 1984ApJ...279..763N, 2006A&A...446.1027C}, while in the turbulent dynamo, magnetic fields result from the interaction of flow turbulence \citep{1993SoPh..145..207D,1996ApJ...469..828D}.
The strength of stellar activities can be traced by X-ray and radio emissions from stellar corona, spectroscopic emission lines (e.g., Ca II H\&K, H$\alpha$) from the chromosphere, and optical flares and cool spots from the photosphere, etc.

Most binaries follow the activity-period relation and the relations between different activity proxies established by single stars \citep{1991A&A...251..183S, 2011ApJ...743...48W}.
This implies the activity properties of binaries are governed by the same physical parameters as single stars, including rotation rates, masses, and ages, which may simply be a consequence that the activities of binaries are dominated by the more active component.
On the other hand, in close binaries, such as RS Canum Venaticorum (RS CVn), BY Draconis (BY Dra), W Ursae Majoris (W UMa), and Algol binary systems, tidal interaction and angular momentum transfer lead to high rotation rates and thus strong magnetic activities of the components.
However, many stars in these binaries, especially RS CVn type binaries, are significantly more active than expected from their rotation rates, called overactive  \citep{1987A&A...177..131R}. 
Possible magnetic interactions between the components may help explain the overactivity, with some observational evidences including extended coronal X-ray emission and the preferential emergence of flux tubes and hot spots on facing hemispheres \citep{2016AAS...22714512H}.

Therefore, (close) binaries become excellent tracers of the upper limits of stellar activity and can be used to study the mechanism of super-saturation in the activity-period relation, which was explained by convective updrafts or coronal stripping.
In the first scenario, nonuniform heating of the convective envelope leads to a poleward migration of active regions, reducing the filling factor of active regions on stellar surface \citep{1984A&A...133..117V,1997A&A...324..943S,1997A&A...325.1039S,2001A&A...370..157S}, while in the second scenario, the fast rotation of a star results in strong centrifugal forces stripping away the outer layers of the corona, reducing the X-ray emitting volume \citep{1999A&A...346..883J,2000MNRAS.318.1217J}.
Furthermore, binaries can also be used to study the impact of tidal forces on magnetic dynamo.
For example, the tidal force may lead to a 1:1 resonant excitation of the oscillation of the $\alpha$-effect, which is capable of exciting the underlying dynamo \citep{2016SoPh..291.2197S,2019SoPh..294...60S,2023SoPh..298...90K}, although this remains a topic of debate \citep{2005SoPh..229..175D,2022SoPh..297..107N,2023A&A...671A..87W}.
On the other hand, tidal effects on rising flux tubes can lead to the formation of clusters of flux tube eruptions at preferred longitudes \citep{2003A&A...405..291H,2003A&A...405..303H}.
Besides, binaries exhibit scaled-up interactions that may also occur in a star-planet system, providing valuable insights into the habitability of exoplanets.

Recent ground-based and space-borne photometric sky surveys, such as ASAS-SN \citep{2017PASP..129j4502K}, ZTF \citep{2019PASP..131a8002B}, WISE \citep{2010AJ....140.1868W}, Kepler \citep{2010Sci...327..977B}, TESS \citep{2015JATIS...1a4003R}, and Gaia
\citep{2016A&A...595A...1G}, have provided large numbers of time-series observations of rotational modulators, offering us a great opportunity to investigate stellar magnetic activity and corresponding dynamo process in binaries \citep[e.g.,][]{2022MNRAS.514.4932C}.
In this study, we choose to utilize the spectroscopic data from the Large Sky Area Multi-Object Fiber Spectroscopic Telescope (LAMOST), together with those photometric surveys, to study the activity variation from single stars to binary systems, across a range of stellar rotation rates and temperatures (i.e., masses).
This approach will enable us to understand the impact of tidal and magnetic interactions on stellar magnetic activity more comprehensively.
The structure of this paper is organized as follows: Section 2 details our sample selection and data reduction process; Section 3 presents the result and discussion; followed is a short summary in Section 4.

\section{Targets selection and Methods}

\subsection{Sample construction}

LAMOST is a 4 m quasi-meridian reflecting Schmidt telescope \citep{2012RAA....12.1197C, 2012RAA....12..723Z} designed with a wide field of view for astronomical spectroscopic survey.
It is renowned for its ability to obtain more than 3000 spectra in a single exposure with a limiting magnitude as faint as $r =$ 19 mag at the low-resolution. 
It performs both low- and medium-resolution spectral observations with $R \sim$ 1800 and $\sim$ 7500, respectively. 
The low-resolution spectrum covers a wavelength range from 3690 \AA \ to 9100 \AA \ \citep{2015RAA....15.1095L}, while the medium-resolution spectrum comprises a blue band from 4950 \AA \ to 5350 \AA \ and a red band from 6300 \AA \ to 6800 \AA \ \citep{2021RAA....21..292W}.

First, we used the machine-learning algorithm Uniform Manifold Approximation and Projection \citep{McInnes2018UMAPUM} to select H$\alpha$ emission spectra from the 
LAMOST DR7 low-resolution database \citep[see][for more details]{2021ApJS..257...65S}.
This leads to 170,383 spectra showing clear stellar activities.
By a further constraint of signal-to-noise ratio (SNR) with SNR$_g$ $>$ 20, we finally derived 18,382 spectra.
Second, we excluded Young Stellar Objects (YSOs) from our sample by cross-matching the YSO catalogs from \citet{2016MNRAS.458.3479M,2019MNRAS.487.2522M,2023A&A...674A..21M} and \cite{2022yCat.1358....0G}.
Third, we removed subgiants and giants by using the absolute magnitude of $M_G < 4.5$ mag. In this step, we derived the distance measurements from Gaia EDR3 \citep{2021AJ....161..147B}, and removed the objects with distances larger than 5 kpc and relative parallax uncertainties larger than 0.2.
In addition, the reddening $E(B-V)$ was derived from the Pan-STARRS DR1 (PS1) 3D dust map  with $E(B-V) =0.884 \times$ Bayestar19, the latter being a measure of extinction given by \cite{2019ApJ...887...93G}.
For sources without extinction estimation from the PS1 dust map, we used the SFD dust map \citep{1998ApJ...500..525S} with $E(B-V) =0.884 \times E(B-V)_{\rm SFD}$ as a complement and only kept the sources with $E(B-V)$ $<$ 0.1.

After the final manual check of the spectra and the exclusion of bad ones, our dataset includes 7,836 individual stars with a total 8,997 spectra.

\subsection{Calculation of $S$ index}

\begin{figure}[ht!]
\centering
   \includegraphics[width=0.45\textwidth]{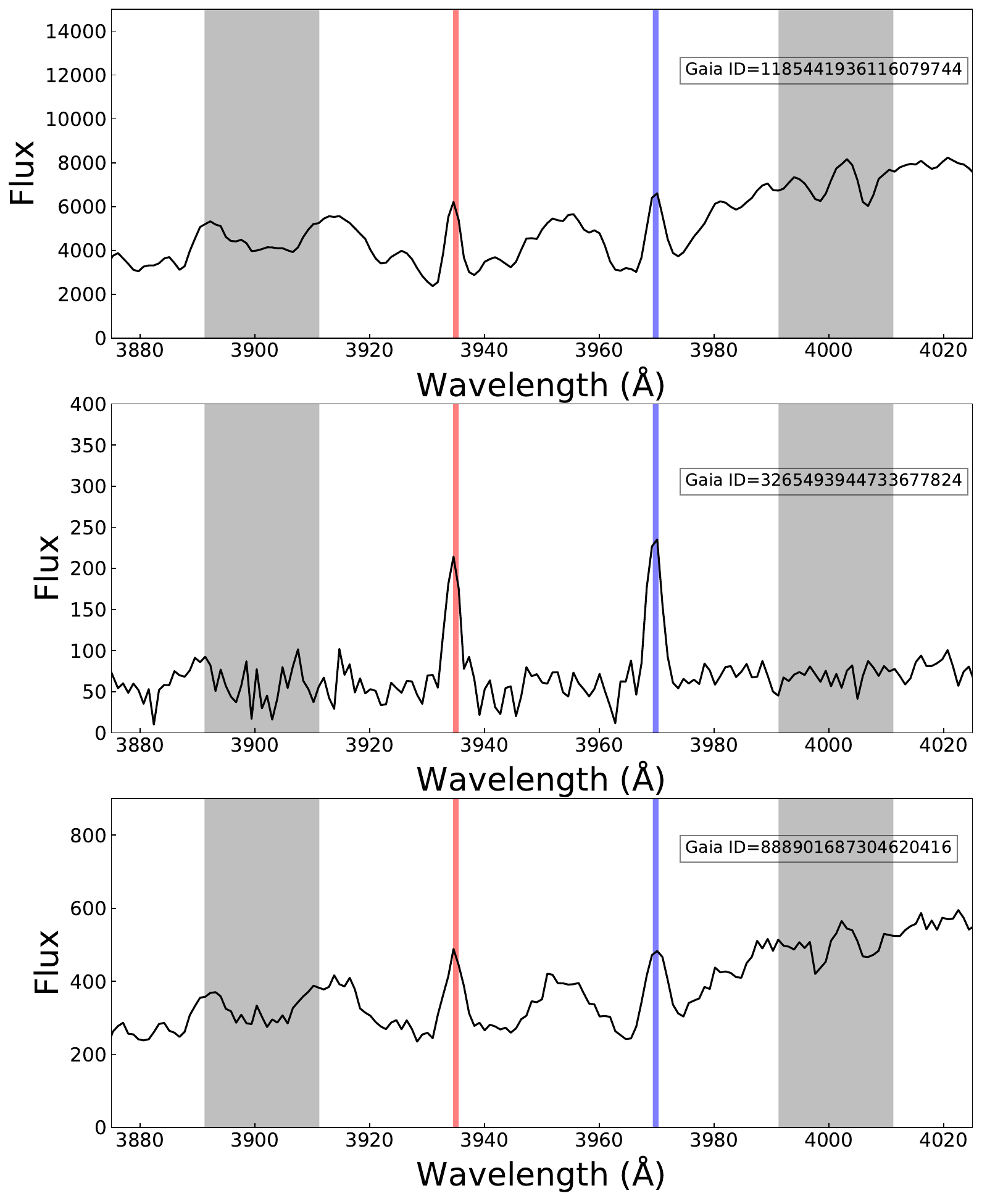}
    %    \plotone{fig/Figure4a.pdf}
%	\plotone{fig/Figure4b.pdf}
	\caption{Examples of the LAMOST low-resolution spectra for our sample stars. The red and green lines mark the Ca K and H lines, while the shaded regions represent the spectral range for continuum estimation.
\label{fig:88}}
\end{figure}

We utilized LAMOST spectra to calculate the canonical $S$ index  \citep{1978PASP...90..267V} to describe the emission in the Ca II H and K lines, which is defined as:
\begin{equation}
S= 8 \alpha \cdot\frac{H+K}{R+V}.
\end{equation} 
The $H$ and $K$ values represent the total fluxes within the core regions of the Ca II H\&K spectral lines. These fluxes are calculated using a triangular integration function, with a Full Width at Half Maximum (FWHM) of 1.09 \AA, centered specifically at wavelengths of 3968 \AA\ and 3934 \AA\ for H and K lines, respectively(Figure \ref{fig:88}). 
The parameters $R$ and $V$ correspond to the integrated fluxes within the adjacent pseudo-continuum regions employing a rectangular integration function featuring a width of 20 \AA\ and centered at 4001 \AA\ and 3901 \AA, respectively(Figure \ref{fig:88}).
We adopt a value of 1.8 for $\alpha$ with the LAMOST LRS data as suggested by \citet{2016NatCo...711058K}. 

\subsection{Stellar rotation periods, orbital periods, and variable types}
\label{sec:catalog}

A number of catalogs, from various photometric/spectroscopic/astrometric sky survey, have provided rotation periods of single stars and orbital periods of binaries, including
Kepler \citep[e.g.,][]{2013MNRAS.432.1203M,2014ApJS..211...24M, 2016AJ....151...68K,2019ApJS..244...21S}, K2 \citep{2020A&A...635A..43R}, ZTF \citep{2020ApJS..249...18C}, ASAS-SN \citep[e.g.,][]{2019MNRAS.486.1907J}, Catalina \citep{2014ApJS..213....9D}, WISE \citep{2018ApJS..237...28C}, 
Gaia \citep{2023A&A...674A..14R}, Gaia DR3 {\sc nss\_two\_body\_orbit} catalog \citep{2023A&A...674A..34G}, TESS \citep[e.g.,][]{2022RNAAS...6...96H, 2022ApJS..258...16P}, GCVS \citep{2017ARep...61...80S}, OGLE \citep{2016AcA....66..405S}, LAMOST DR7 \citep{2022A&A...660A..38W}, MEarth \citep{2016ApJ...821...93N}, and HATNet \citep{2011AJ....141..166H}.
We cross-matched our sample with these catalogs, using a matching radius of 3$\arcsec$, to derive the periods, as well as their stellar variable types.
For sources observed by multiple surveys, we selected the periods and variable types based on the priority sequence as mentioned above. This led to 817 rotational single stars and 788 binaries, the latter of which includes 216 eclipsing stars, 28 RS CVn stars, and 544 BY Dra stars.

\section{Results and Discussions}

\begin{figure}[ht!]
 \center
   \includegraphics[width=0.48\textwidth]{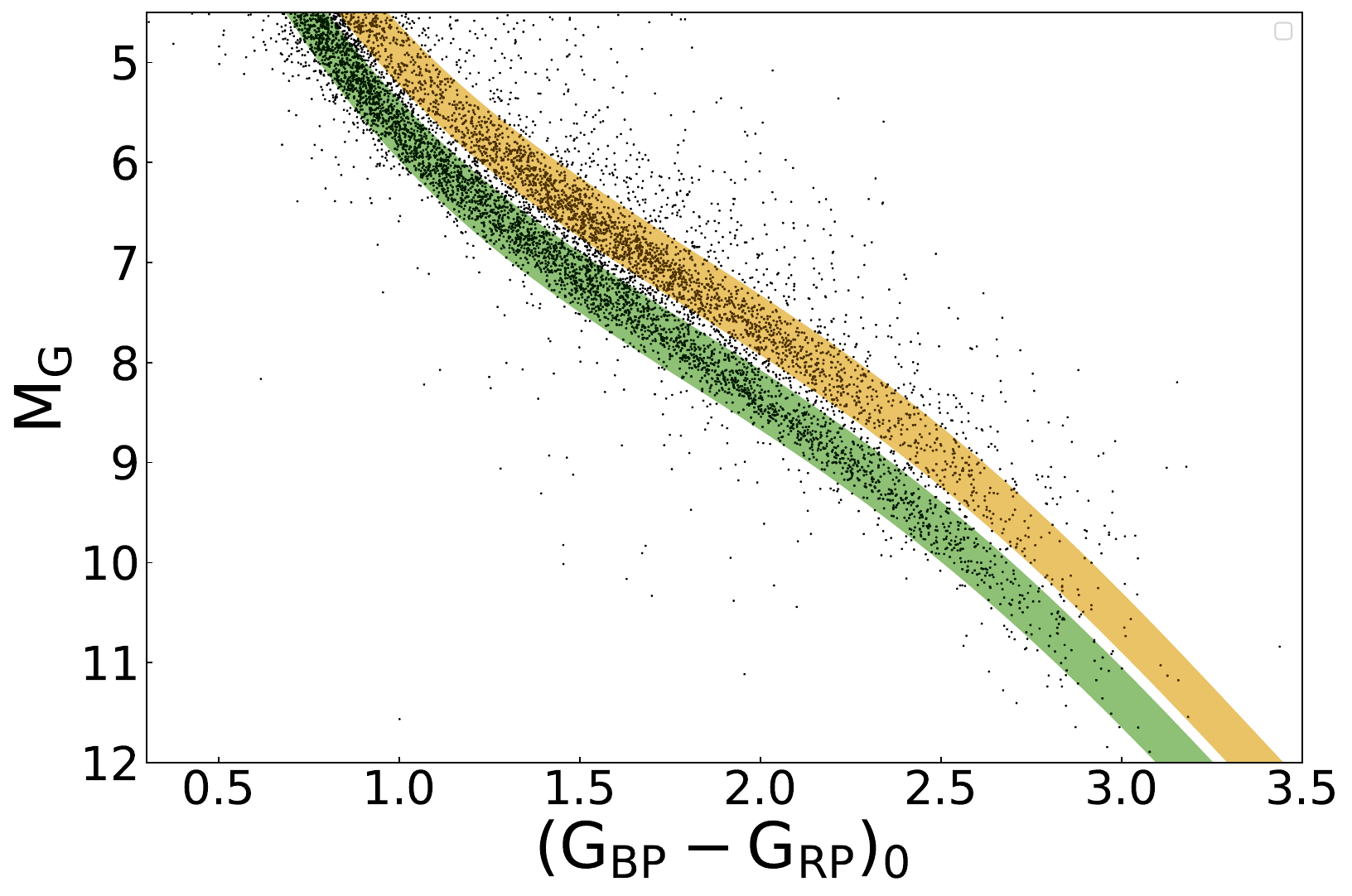}
   \caption{ Hertzsprung-Russell diagram of our sample stars. The binary and single-star populations are marked with yellow and green slices, respectively.}
 \label{fig:S}
 %\caption{Distribution of the four sliced stars in CMD.\label{fig:S}}
\end{figure}

In this section, we aim to compare the stellar activity between single and binary stars.
In the Hertzsprung-Russell (HR) diagram (Figure \ref{fig:S}), the binary belt is clearly visible at 0.75 magnitudes above the main sequence, corresponding to unresolved binaries containing two identical stars, observed with the same color but twice the luminosity of an equivalent single star \citep{2018A&A...616A..10G}.
First, we performed a polynomial fitting to the main sequence and derived the equation
\begin{equation}
\begin{split}
M_{\rm G} = & 0.2339+7.0149\ (G_{\rm BP}-G_{\rm RP}) \\
&-2.1773\ (G_{\rm BP}-G_{\rm RP})^2\\
&+0.6934\ (G_{\rm BP}-G_{\rm RP})^3\\
&-0.07251\ (G_{\rm BP}-G_{\rm RP})^4.
\end{split}
\label{equ:split}
\end{equation} 
Second, we selected a binary population and a single-star population from the sample using two slices following the trend of Equation \ref{equ:split}.
These two populations mainly correspond to the binary belt and the main sequence, respectively.
The height of each slice (i.e., the brightness range for the same colour) was set to be 0.6 mag. 
Between the two slices, we removed a region of 0.15 magnitudes (in height) in order to enhance the distinction between them.
This led to 2433 and 3637 stars in the binary and single-star slices, respectively. 
Although there may be binaries, consisting of a brighter and a fainter star, or triple systems randomly distributed in the HR diagram, the binary fraction in the binary slice is expected to be much higher than that in the single-star slice.

\begin{table*}[!ht]
\centering
\setlength{\tabcolsep}{1pt}
\begin{threeparttable}
\caption{Stellar parameters of the sample stars. \label{Table1}}
    \begin{tabular}{rcccccccc}
    \hline
Gaia ID & R.A. & Decl. & $S$ index & $G_{\rm BP} - G_{\rm RP}$ & $M_G$ & Flag$^a$ & Period & Database   \\ 
 & (deg) & (deg) &  & (mag) & (mag) &  & (day) &    \\
\hline
120001145732321536 & 54.24082  & 29.28797  & 0.49  & 1.01  & 5.29  & ~  \\ 
219244031627479296 & 61.29465  & 35.95703  & 0.14  & 0.66  & 4.62  & ~  \\ 
671703305656347136 & 116.40658 & 19.07022  & 0.96  & 1.56  & 7.37  & ~  \\ 
710247613482721536 & 129.89198 & 32.86587  & 0.62  & 1.43  & 6.18  & b  \\ 
1783025848682256768 & 332.34528 & 22.97542  & 0.47  & 1.08  & 6.04  & ~  \\ 
1894578179565911296 & 336.72892 & 30.17105  & 2.54  & 1.78  & 7.96 & & 306.25534 & ASAS-SN  
\\ 
2065940880688308480 & 313.83397 & 42.56631  & 0.24  & 0.82  & 5.04  & s  \\ 
2067307023883242240 & 307.39033 & 39.84348  & 0.22  & 1.00  & 5.80  & s  \\ 
2067556819183564928 & 306.04291 & 40.80641  & 0.14  & 0.87  & 4.86  & s  \\ 
2067663540529838592 & 306.00315 & 40.89280  & 0.22  & 1.03  & 5.94  & s  \\ 
2069756392191707392 & 308.15209 & 43.09314  & 0.18  & 1.10  & 5.38  & b  \\ 
2797676726644861952 & 3.19585   & 19.50399  & 0.35  & 0.94  & 5.37  & ~  \\ 
3019756802483967232 & 92.21274  & -6.57617  & 0.19  & 0.83  & 4.86  & s  \\ 
3023162715844336128 & 85.95600  & -5.00640  & 0.65  & 1.06  & 5.23  & b  \\ 
3210609210495286016 & 81.55687  & -4.16286  & 0.47  & 1.01  & 4.62  & & 0.82689 & ASAS-SN
\\ 
3216541728561016704 & 85.02470  & -2.05755  & 1.40  & 1.01  & 4.53  & & 2.39601 & ASAS-SN 
\\ 
3327243101868702336 & 97.99821  & 9.82066   & 0.87  & 1.49  & 7.18  & ~  \\ 
3377048603487213440 & 94.71393  & 22.79092  & 0.15  & 0.96  & 5.05  & b  \\
3414109846220439680 & 77.80178  & 20.90110  & 1.64  & 1.83  & 8.01  & ~  \\ 
3994816465752179968 & 165.76459 & 22.88967  & 1.23  & 1.72  & 6.50  & ~  \\ 
... & ... & ... & ... & ... & ... & ... & ... & ...  \\
        \hline 
    \end{tabular}
    \begin{tablenotes}
\item[$a$] The flag ``b" indicates the star belongs to the binary population (i.e., binary belt), while the flag ``s" indicates the star belongs to the single-star population (i.e., main sequence).
\end{tablenotes}
This table is available in its entirety in machine-readable and Virtual Observatory (VO) forms in the online journal. A portion is shown here for guidance regarding its form and content.
\end{threeparttable}
\end{table*}

\subsection{Distribution of $S$ index for single stars and binaries}
\label{dis.sec}

%Figure \ref{fig:S} (right panel) presents the distribution of $S$ index for different subsamples. It clearly shows that subsamples $a$ and $b$, which have a larger proportion of binary stars, exhibit higher $S$ index compared to subsamples $c$ and $d$.
%
Given the dependence of $S$ index on stellar types \citep{1980PASP...92..385V,2013A&A...549A.117M,2018A&A...616A.108B}, 
a comparison of $S$ index between the two subsamples (``b" for binaries and ``s" for single stars) within the same narrow colour range is necessary.
Therefore, we divided each subsample into 10 bins using the colour $G_{\rm BP}-G_{\rm RP}$, ranging from 1.0 to 3.0 with a step of 0.2.
Figure \ref{fig:S3} displays the distribution of the $S$ index for the two subsamples in each colour bin, and 
it can be seen that in the color range of 1.0--1.8, mainly consisting of K-type stars, binaries show a higher level of S index compared to single stars.
However, when $G_{\rm BP}-G_{\rm RP} > 1.8$, their activity levels become indistinguishable.

\begin{figure*}[ht!]
\centering
   \includegraphics[width=0.48\textwidth]{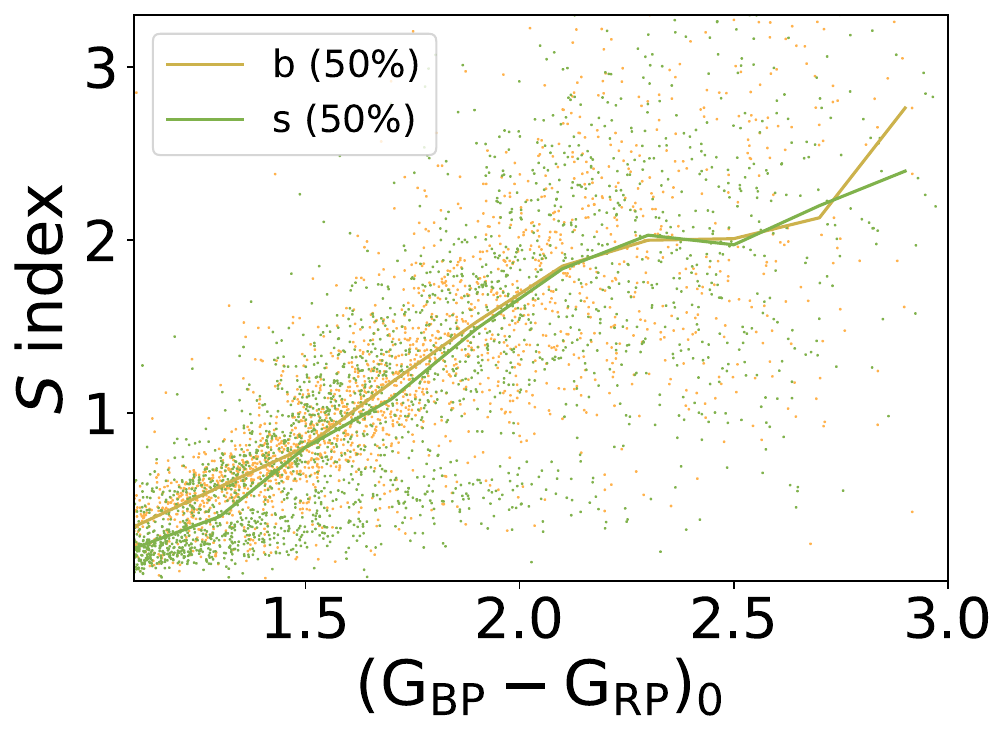}
  \includegraphics[width=0.48\textwidth]{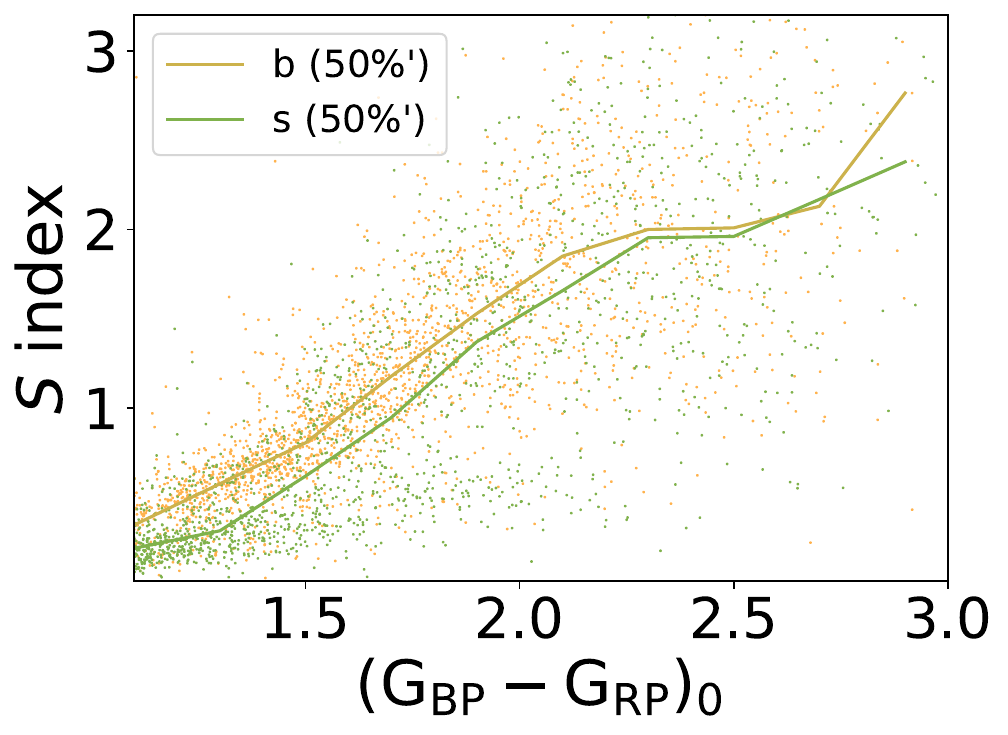}
    %    \plotone{fig/Figure4a.pdf}
%	\plotone{fig/Figure4b.pdf}
	\caption{$S$ index versus stellar colour. Left panel: The yellow and green lines represent the median values (Column 50\% in Table \ref{Table2}) of the $S$ indices in each colour bin for the binary and single-star samples, respectively. The yellow and green dots correspond to the $S$ indices for the binary and single-star samples. Right panel: The yellow and green lines represent the median values (Column 50\%' in Table \ref{Table2}) of the $S$ indices in each colour bin for the binary and single-star samples, respectively. 
\label{fig:S2}}
\end{figure*}

Table \ref{Table2} provides the 50$th$ and 90$th$ percentiles for these distributions.
Furthermore, to enhance clarity in the comparison, for the single-star subsample (i.e., subsample ``s"), we excluded binaries classified by the catalogs in Section \ref{sec:catalog}. 
Objects with radial velocity variation larger than 10 km/s were also excluded.
The corresponding 50$th$ and 90$th$ percentiles of the new $S$-index distributions are flagged as 50\%' and 90\%'.
Figure \ref{fig:S2} displays the relationship between $S$ index and stellar colour, showing that binaries have a higher level of $S$ index compared to single stars when the $G_{\rm BP}-G_{\rm RP}$ color is bluer than $\approx$1.5.
More notably, after removing binaries from the single-star slice, the binary belt show a significantly higher level of $S$ index in the color range from 1.0 to 2.5 (Figure \ref{fig:S2} right panel).

\subsection{Distribution of rotation periods, and orbital periods}
\label{sec:period}

To explore the cause of the difference in the $S$-index distribution of binaries and single stars, we examined the distribution of orbital and rotation periods.
In Section \ref{sec:catalog}, we cross-matched our total sample with some variable catalogs to derive orbital periods for binaries and rotational periods for single stars.
However, the binaries and single stars in these catalogs may not be perfectly aligned with the binary and main-sequence slices.
This can be due to uncertainties of magnitudes or parallaxes, or misclassifications in these catalogs.
To refine the sample, we further cross-matched the binary and single-star slices with the binary sample with orbital periods and the single-star sample with rotational periods.
Finally, we analyzed the distribution of orbital periods from 315 binaries in the binary slice and rotational periods from 327 single stars in the main-sequence slice.

\begin{figure*}
%	\plotone{fig/Figure5.pdf}
\centering
  \includegraphics[width=0.95\textwidth]{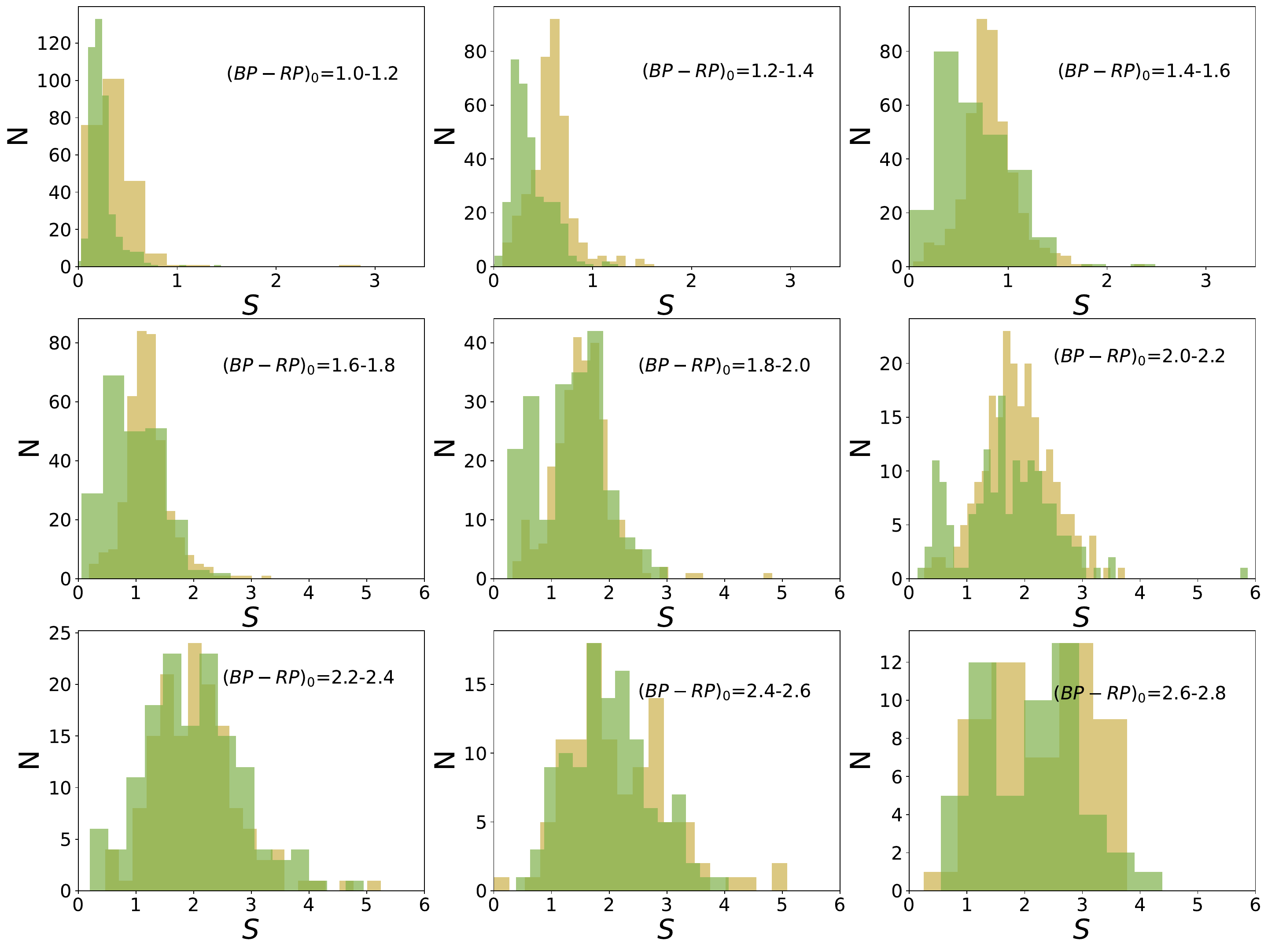}
 \caption{Distributions of the $S$ indices of the two samples in each $G_{\rm BP}-G_{\rm RP}$ bin. The yellow and green histograms represent the binary and single-star samples, respectively.
 \label{fig:S3}}
\end{figure*}

\begin{figure*}[ht!]
\centering
   \includegraphics[width=0.98\textwidth]{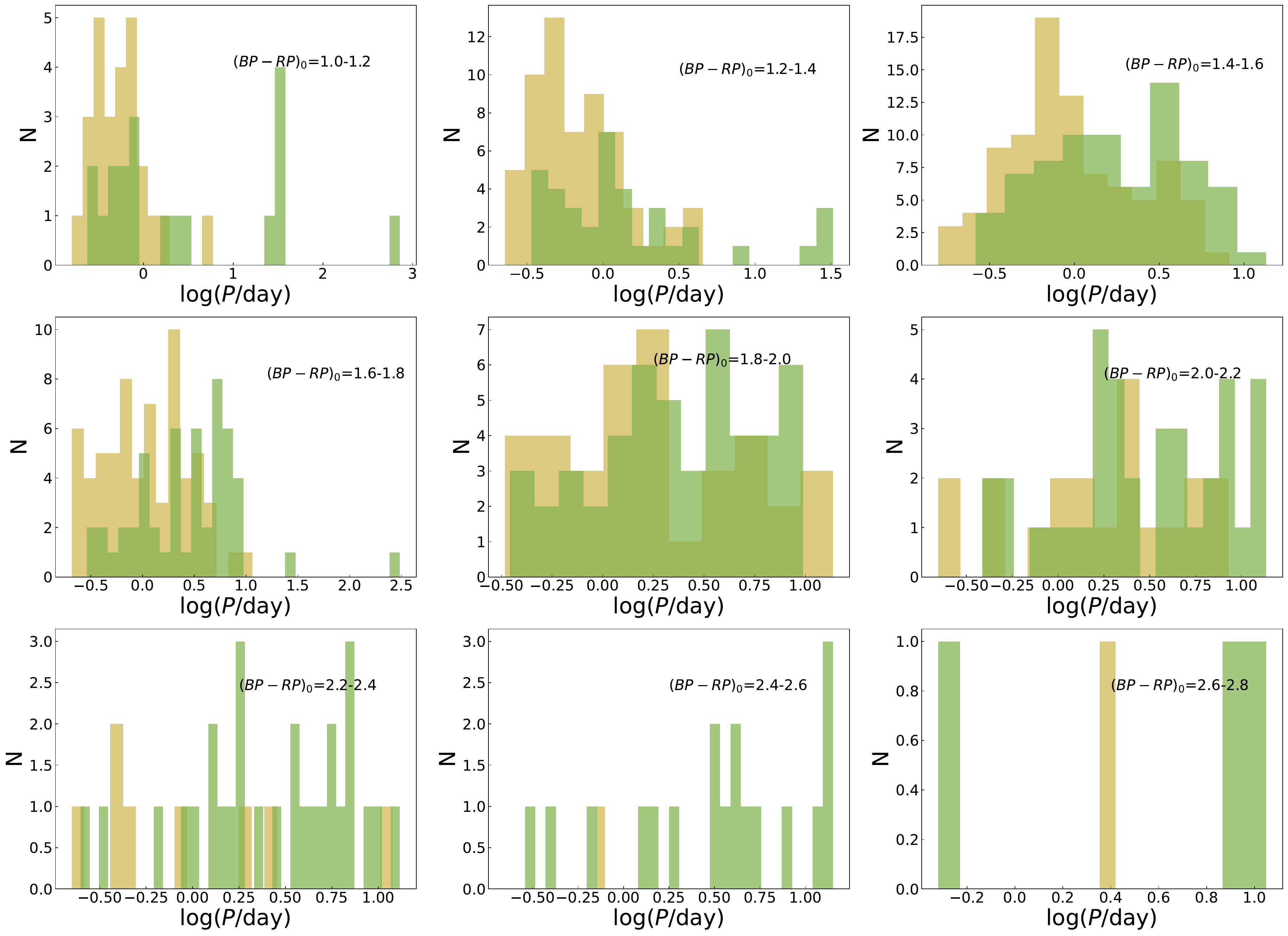}
    %    \plotone{fig/Figure4a.pdf}
%	\plotone{fig/Figure4b.pdf}
	\caption{Distributions of rotation periods for single stars and orbital periods for binaries in each $G_{\rm BP}-G_{\rm RP}$ bin. The yellow and green histograms represent the binary and single-star samples, respectively. 
\label{fig:per}}
\end{figure*}

Most binaries in our sample have an orbital period less than $\approx$20 days (Figure \ref{fig:per}), suggesting the components have been synchronized and the spin period of one or both components equals the orbital period. 
The period distribution shows that, in the color range of 1.0--1.8, binaries have shorter orbital periods compared to rotational periods of single stars.
This helps explain the distribution of $S$ index in Figure \ref{fig:S3}.
%
%The pattern of activity across the HR diagram for binary stars are more complex than single stars, for which the activity is mainly the result of stellar evolution and magnetic braking.
%
For binaries, the tidal interaction plays an important role in influencing stellar evolution through the transfer of angular momentum \citep{2021Univ....7..440D}.
On the one hand, orbital synchronization in close binaries can effectively increase stellar spin rate and thus suppress the magnetic braking effect and enhance stellar activity.
On the other hand, tidal distortion and magnetic interaction between the components lead to preferential emergence of flux tubes \citep{2003A&A...405..291H,2003A&A...405..303H}, hot spots on facing hemispheres, and extended coronal X-ray emission in RS CVn binaries \citep{1996ApJ...473..470S}, perhaps lasting over a decade.
For example, a significant increase in spot coverage was found on the hemisphere facing the white dwarf component in binaries like V426 Oph, SS Cyg, BV Cen, and AE Aqr \citep{2016AAS...22714512H}.

\subsection{Activity-period relation for single stars and binaries}

\begin{figure*}[ht!]
\centering
%	\plotone{fig/Figure6.pdf}
 \includegraphics[width=0.97\textwidth]{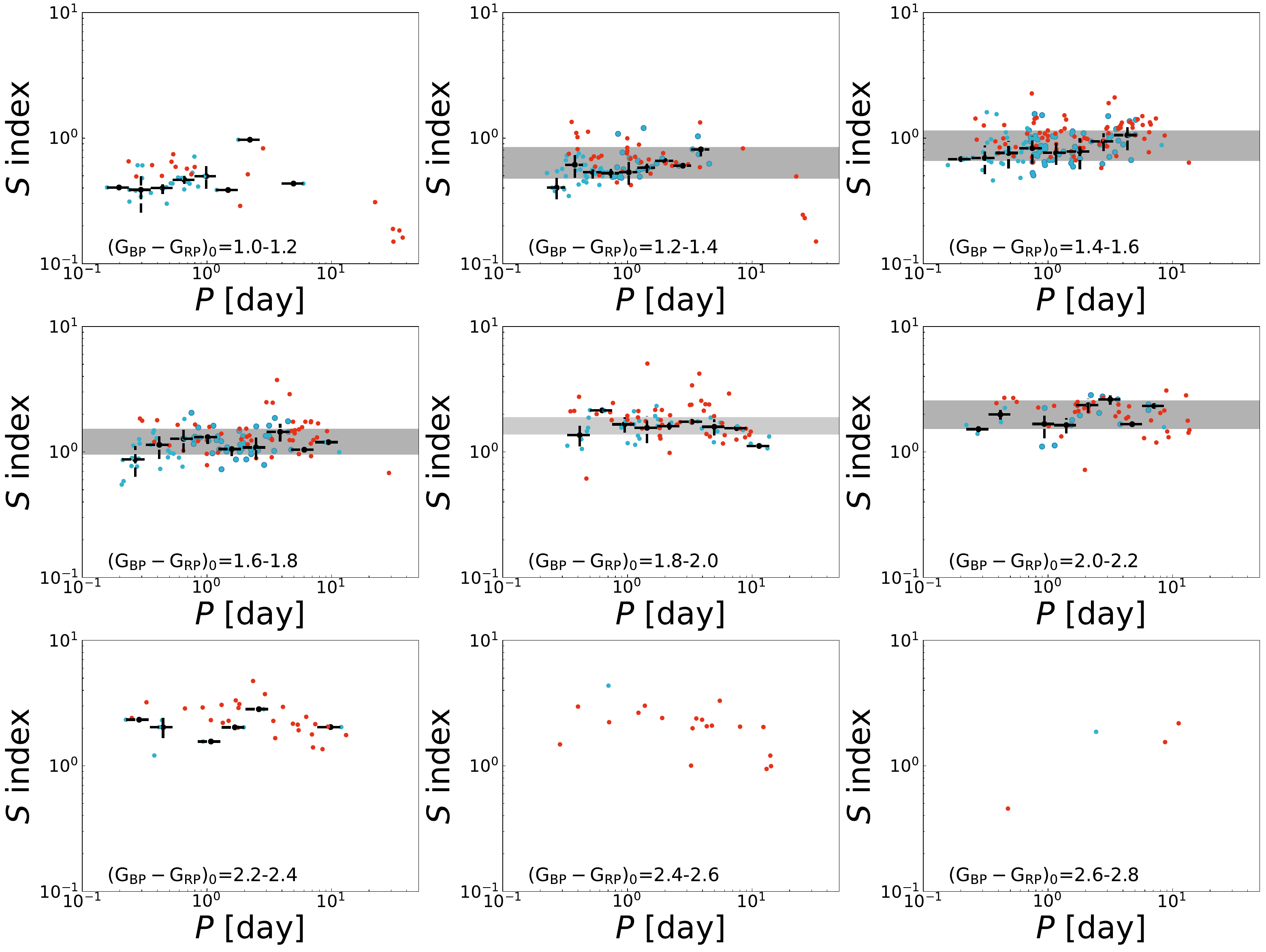}
 \caption{$S$ indices versus rotation periods for single stars (red dots) and orbital periods for binaries. The pluses represent the median $S$ index value in each period bin, and the uncertainties are calculated using the 16{\it th} and 84{\it th} percentiles. Shaded region indicates the 16{\it th} and 84{\it th} percentiles of $S$ indices for binaries with periods ranging from 0.7 to 7 days. \label{fig:S6}}
 \label{fig:P}
\end{figure*}

Figure \ref{fig:P} shows the activity-period relation for the sample stars. 
In each color bin, both single stars and binaries fall on a common sequence in the relation $S$ index versus period.
This indicates stellar activities of binaries are mainly governed by the more active component \citep{Schrijver_Zwaan_2000}.
For K1--K5 type stars ($G_{\rm BP}-G_{\rm RP} < 1.4$), both the saturated regime  (i.e., with high and constant activity levels) and the unsaturated regime  (i.e., with declining activity levels and periods longer than $\approx$10 days) are observable, while for K6--M4 type stars ($G_{\rm BP}-G_{\rm RP} > 1.4$), only the saturated region is evident.
The saturation value varies each color bin, ranging from 0.5 ($G_{\rm BP}-G_{\rm RP} = 1.0$--1.2) to 2.6 ($G_{\rm BP}-G_{\rm RP} = 2.6$--2.8).
This variation can be attributed to the thickening of the convective envelope as $T_{\rm eff}$ decreases, leading to a clear increase in the $S$ index.

Interestingly, in some color bins (Figure \ref{fig:P}), especially $G_{\rm BP}-G_{\rm RP} =$ 1.2--2.2, there is a slight decline in the saturation limit for binaries with very short orbital periods. 
%\textcolor{red}{We further focus on investigating the phenomenon of $S$ index decrease specifically in binaries with very short orbital periods, while such a decrease is not observed in single stars.}
%
We calculated the significance with the following steps. By using the binaries with periods ranging from 0.7 to 7 days, we computed the 50{\it th} percentile as the median value and half of the (84{\it th} $-$ 16{\it th}) percentiles as the standard deviation $\sigma$. Then we calculated the difference between the leftmost black plus and the median value in each panel. We found in all color bins, the significance (i.e., the ratio of the difference to the standard deviation) is approximately 1$\sigma$.
Although the significance of the declining trend may not be high, we would still like to provide a brief discussion on the potential mechanisms.

One explanation is the the decline trend is caused by reduced area of active regions, similar to the phenomenon of super-saturation \citep{1996AJ....112.1570P}.
The super-saturation behaviour has been only observed in the X-ray band \citep[e.g.][]{2011ApJ...743...48W}, and its physical origin remains debated.
Corresponding theories include decreased filling factor of coronal active regions on the stellar surface \citep{1997A&A...325.1039S,2001A&A...370..157S} or reduced X-ray emitting volume due to centrifugal stripping  \citep{1999A&A...346..883J, 2000MNRAS.318.1217J}.
In previous studies, the lack of observed super-saturation in chromospheric emission \citep{2007ApJ...666..393C,2008ApJ...687.1264M,2009MNRAS.399..888M,2010MNRAS.407..465J} was used to argue in favor of the coronal stripping scenario, since the scenario of the migration of active regions towards the poles could cause super-saturation in chromospheric emission as well.
Although we also observe no reduction in the $S$ index for single stars, the binaries in our sample suggest evidence of possible super-saturation in chromospheric emission.
Recently, \citet{2022MNRAS.514.4932C} found a decline in the saturation value of $S_{\rm ph}$ index for BY Dra-type variables with short periods, most of which are also binaries.
Although these authors explained that it was caused by bright faculae partially canceling out dark spots or by the method used to estimate the $S_{\rm ph}$ index, it is also possible that the super-saturation phenomenon is at play.
In this case, the super-saturation observed in X-ray, chromospheric, and photospheric activity proxies suggests the decrease in the filling factor of active regions is more likely the mechanism for super-saturation.

Another explanation is that the strong tidal forces may suppress the magnetic activity in very close binaries. 
As shown in Section \ref{dis.sec}, binaries have enhanced activity compared to single stars due to the increase in spin rate caused by orbital synchronization.
However, for very close binary, a more complex situation should be considered.
For example, the differential rotation on the surface of the binaries might be suppressed by the tidal forces.
Moreover, tidal interactions are expected to induce large-scale 3D shear and/or helical flows in stellar interiors that can notably perturb the stellar dynamo \citep{2015IAUS..307..330A}.
These processes may strongly affect the stellar dynamo and finally the reduce stellar activities.

%\begin{figure*}[ht!]
%\centering
%   \includegraphics[width=0.50\textwidth]{fig/Figure7.pdf}
%  
%    %    \plotone{fig/Figure4a.pdf}
%%	\plotone{fig/Figure4b.pdf}
%	\caption{Distribution of the binary periods.
%\label{fig:78}}
%\end{figure*}

\begin{table}[t!]
\centering
\resizebox{\columnwidth}{!}{
%\small
\setlength{\tabcolsep}{0.8pt}
\begin{threeparttable}
\caption{Distributions of the $S$ indices of the two sliced populations in each $G_{\rm BP}-G_{\rm RP}$ bin. \label{Table2}}
\begin{tabular}{lcccccccc}
\hline
$G_{\rm BP}-G_{\rm RP}$ & Spec. Type & Flag$^a$ & 50$\%^b$ & 90$\%^b$& N & 50$\%$'$^b$ & 90$\%$'$^b$ & N' \\
\hline
1.0-1.2 & K1-K3 & b & 0.35 & 0.58 & 234 & 0.35 & 0.58 & 234 \\ 
 & & s & 0.22 & 0.48 & 515 & 0.21 & 0.39 & 447 \\ 
\hline
1.2-1.4 & K3-K5 & b & 0.58 & 0.78 & 361 & 0.58 & 0.78 & 361 \\ 
  & & s & 0.40 & 0.76 & 447 & 0.31 & 0.63 & 323 \\ 
\hline
1.4-1.6 & K5-K6.5 & b & 0.80 & 1.13 & 435 & 0.80 & 1.13 & 435 \\ 
 & & s & 0.80 & 1.25 & 445 & 0.62 & 1.12 & 260 \\ 
\hline
1.6-1.8 & K6.5-K9 & b & 1.18 & 1.65 & 387 & 1.18 & 1.65 & 387 \\ 
 & & s & 1.08 & 1.72 & 352 & 0.95 & 1.56 & 225 \\ 
\hline
1.8-2.0 & K9-M1 & b & 1.53 & 2.10 & 280 & 1.53 & 2.10 & 280 \\ 
 & & s & 1.49 & 2.21 & 314 & 1.37 & 2.01 & 204 \\ 
\hline
2.0-2.2 & M1-M2 & b & 1.85 & 2.66 & 223 & 1.85 & 2.66 & 223 \\ 
 & & s & 1.83 & 2.64 & 236 & 1.66 & 2.66 & 162 \\ 
\hline
2.2-2.4 & M2-M2.5 & b & 2.00 & 2.95 & 149 & 2.00 & 2.959 & 149 \\ 
 & & s & 2.03 & 3.18 & 184 & 1.95 & 3.00 & 142 \\ 
\hline
2.4-2.6 & M2.5-M3.1 & b & 2.01 & 3.27 & 107 & 2.01 & 3.27 & 107 \\ 
 & & s & 2.00 & 3.04 & 144 & 2.00 & 3.07 & 113 \\ 
\hline
2.6-2.8 & M3.1-M3.5 & b & 2.13 & 3.31 & 51 & 2.13 & 3.31 & 51 
\\ 
 & & s & 2.20 & 3.32 & 66 & 2.17 & 3.16 & 52 \\ 
\hline
2.8-3.0 & M3.5-M4 & b & 2.76 & 3.65 & 18 & 2.76 & 3.65 & 18 
\\ 
 & & s & 2.40 & 3.44 & 28 & 2.38 & 3.27 & 27 \\ 
\hline
total &  & b & 0.95 & 2.19 & 2433 & 0.95 & 2.19 & 2433  \\
 & & s  & 0.50 & 2.05 & 3637 & 0.31 & 1.93 & 2819 \\
 \hline
 \end{tabular}
    \begin{tablenotes}
    \item[$a$] The flag ``b" indicates the star belongs to the binary population (i.e., binary belt), while the flag ``s" indicates the star belongs to the single-star population (i.e., main sequence).
    \item[$b$] The 50\% and 90\% represent the 50$th$ and 90$th$ percentiles of $S$-index distribution in each color bin, respectively.
    \end{tablenotes}
    \end{threeparttable}}
\end{table}

\section{Summary}

In this study, we investigated the chromospheric activities of single stars and binaries by using the low-resolution spectra from the LAMOST DR7 dataset.
In total, our sample includes 7836 stars with 8997 spectra. The $S$ index was calculated using the Ca II H\&K lines for each spectrum.
To compare the activities between single and binary stars, we selected two subsamples following the main sequence trend in the HR diagram.

Our results show that, in the $G_{\rm BP}-G_{\rm RP}$ color range of 1.0--1.8 (i.e., mainly K-type stars), binaries tend to exhibit statistically higher levels of activity compared to single stars in the, which could be attributed to tidal synchronization or magnetic field interactions.
Simultaneously, binaries do not exhibit double the $S$ index of single stars, suggesting stellar activities of binaries are predominantly governed by the more active component.
In the activity-period relation, both single stars and binaries fall on a common sequence, further indicating that the chromospheric activities of binaries are dominated by the more active components.

On the other hand, within certain color ranges, the presence of a super-saturation regime (i.e., a slight decline of the $S$ index for smaller periods) is observed for binary stars. 
The significance of the super-saturation is not substantial (i.e., about 1$\sigma$), thus a larger sample of binaries with more accurate measurements of chromospheric indices is needed to validate this observation.
If the trend is not caused by sample selection effects, this may mark the first observation of super-saturation in chromospheric activity.
The super-saturation observed in X-ray, chromospheric, and photospheric activity proxies in recent studies indicates the decrease in the filling factor of active regions is more likely the mechanism for super-saturation.
Another explanation for the slight decline trend is that strong tidal forces may significantly affect magnetic dynamo and reduce stellar activities in very close binaries, by suppressing the differential rotation and the convection. 

The activity patterns across the HR diagram for close binary stars are more complex than those of single stars, and they are worthy of more detailed future studies, especially with larger binary samples and more accurate activity measurements.

\section*{acknowledgements}
We thank the anonymous referee for helpful comments and suggestions that have significantly improved the paper. 
The Guoshoujing Telescope (the Large Sky Area Multi-Object Fiber Spectroscopic Telescope LAMOST) is a National Major Scientific Project built by the Chinese Academy of Sciences. Funding for the project has been provided by the National Development and Reform Commission. LAMOST is operated and managed by the National Astronomical Observatories, Chinese Academy of Sciences.
Some of the data presented in this paper were obtained from the Mikulski Archive for Space Telescopes (MAST).
This work presents results from the European Space Agency (ESA) space mission {\it Gaia}. {\it Gaia} data are being processed by the {\it Gaia} Data Processing and Analysis Consortium (DPAC). Funding for the DPAC is provided by national institutions, in particular the institutions participating in the {\it Gaia} MultiLateral Agreement (MLA). The {\it Gaia} mission website is https://www.cosmos.esa.int/gaia. The {\it Gaia} archive website is https://archives.esac.esa.int/gaia.
This study is supported by the National Key Basic R$\&$D Program of China No. 2019YFA0405000 and 2019YFA0405504, the National Natural Science Foundation of China under grant No. 12173013, the Science Research Grants from the China Manned Space Project with No. CMS-CSST-2021-A08, the Strategic Priority Program of the Chinese Academy of Sciences undergrant No. XDB41000000, XDB0560000, the project of Hebei provincial department of science and technology under the grant number 226Z7604G, and the Hebei NSF (No. A2021205006). W.C. thanks the Science Research Grants from the China Manned Space Project.

\bibliographystyle{aasjournal}%{yahapj}
\bibliography{main}

\clearpage

\end{CJK*}
\end{document}